\documentclass[prd,showpacs,preprintnumbers,floatfix,superscriptaddress,10pt]{revtex4}
\usepackage{graphicx}
\usepackage{color}
\begin{document}
%%%   New Definitions
\newcommand{\eg}{{\it e.g.}}
\newcommand{\etal}{{\it et. al.}}
\newcommand{\ie}{{\it i.e.}}
\newcommand{\be}{\begin{equation}}
\newcommand{\dd}{\displaystyle}
\newcommand{\ee}{\end{equation}}
\newcommand{\bea}{\begin{eqnarray}}
\newcommand{\eea}{\end{eqnarray}}
\newcommand{\bef}{\begin{figure}}
\newcommand{\eef}{\end{figure}}
\newcommand{\bce}{\begin{center}}
\newcommand{\ece}{\end{center}}
\def\lsim{\mathrel{\rlap{\lower4pt\hbox{\hskip1pt$\sim$}}
    \raise1pt\hbox{$<$}}}         %less than or approx. symbol
\def\gsim{\mathrel{\rlap{\lower4pt\hbox{\hskip1pt$\sim$}}
    \raise1pt\hbox{$>$}}}         %greater than or approx. symbol
\title{Properties of charmonia in a hot equilibrated medium}
\author{Floriana Giannuzzi}
\email{Floriana.Giannuzzi@ba.infn.it}
\affiliation{ I.N.F.N., Sezione di Bari, I-70126 Bari, Italia, 
Universit\`a degli Studi di Bari, I-70126 Bari, Italy}

\author{Massimo Mannarelli}
\email{massimo@ieec.uab.es}
\affiliation{Instituto de Ciencias del Espacio (IEEC/CSIC) Campus Universitat Aut\`onoma de Barcelona, Facultat de Ci\`encies, Torre C5, E-08193 Bellaterra (Barcelona), Catalonia, Spain}


\begin{abstract}
We investigate the properties of  charmonia   in a thermal medium, showing that with increasing temperature the   decay widths of these mesons behave in a non-trivial way.    Our  analysis is based on a potential model with    interaction potential extracted from thermal lattice QCD calculations of the free-energy of a static quark-antiquark pair.  
We find that in the crossover region some decay widths are extremely enhanced. %, which might be interpreted as a signature of the transition from hadronic matter to the deconfined phase.
In particular, at temperatures $T\sim T_c$ the decay widths of the $J/\Psi$ that depend on the value of the wave function at the origin  are enhanced  with respect to the values in  vacuum by about a factor $2$. In the same temperature range the decay width of the process   $\chi_{c J} \to J/\Psi + \gamma$ is enhanced by approximately  a factor $6$ with respect to the value in vacuum.  At higher  temperatures  the charmonia states dissociate and the widths of both decay processes become vanishing small. 

\end{abstract}
\date{\today}
\maketitle
\section{Introduction}
%{\it Introduction.---} 
One of the central planks of the heavy-ion collision program is    to identify and characterize the properties of deconfined quark matter. 
Many valuable probes of the properties of  the  medium produced in heavy-ion collisions are available, that include   jets, electromagnetic signals and heavy  quarkonia states ($Q \bar Q$)~\cite{Abreu:2007kv}. In particular much work has been devoted to understand how the presence  of deconfined quarks and gluons may affect the binding  of quarkonia~\cite{reviews}. 

Whether or not hadrons survive in the quark gluon plasma  and how their decay widths are modified is a long-standing issue~\cite{Hatsuda:1985eb}.
 Since the pioneering work by Matsui and Satz~\cite{Matsui:1986dk} it is known that the screening effect due to a thermal medium induces the  dissociation of quarkonia states. Subsequent analyses of the binding energies of mesonic states as a function of the temperature have been done by various authors using potential models, with an interaction potential extracted from lattice QCD calculations~\cite{Shuryak:2004tx,Wong:2004zr,Mannarelli:2005pz,Mocsy:2007jz}. These analyses show that  quarkonia dissociate at  temperatures close to the critical temperature of QCD, $T_c$.  Analogous results have been obtained by studying the correlation functions of charmonia above deconfinement~\cite{Asakawa:2003re}.
 
In the present paper we study the properties of heavy quarkonia states with increasing temperature. In particular we are interested in determining how the decay widths for the leptonic, hadronic and radiative channels are influenced by the presence of the thermal medium. In order to achieve this result we first determine the radial wave function of some charmonia states and the corresponding binding energies.  In our analysis we  employ the non-relativistic Scrh\"odinger equation (and for comparison  the relativistic Salpeter equation)   with interaction   potential extracted from thermal lattice QCD simulations.  A first analysis of the decay widths of charmonia have been carried out in~\cite{Mocsy:2004bv} by studying the quark correlators above the deconfinment temperature. Instead we employ a wave equation and study the decay widths for any temperature up to the dissociation temperature of the pertinent quarkonium state. 

The remarkable result of our analysis is that some decay widths change drastically close to $T_c$.
In the transition region  we find that  the decay widths of the $J/\Psi$  that depend on the value of the wave function at the origin change by approximately a factor $2$   with respect to the corresponding values in  vacuum. We also investigate the properties of the $\chi_c$ meson and in particular of the  $\chi_{c J} \to J/\Psi + \gamma$ transition. This  radiative decay contributes to  the total inclusive $J/\Psi$ production by a fraction of about  $0.3$, as determined in proton proton and $\pi^+$ $\pi^-$  data~\cite{Antoniazzi:1992iv}.   We find that this $J/\Psi$ production mechanism is enhanced close to the critical temperature by approximately a factor $6$  with respect to the value in  vacuum.

Although it has been shown in perturbation theory that the potential has a sizable imaginary part~\cite{Laine:2006ns}, we shall employ a real potential as extracted from  lattice QCD calculations of the free-energy.    Since a  non perturbative  estimate of the imaginary part of the potential is not yet available,  
 it is hard to estimate how large will be its contribution  to the processes we are considering. 

\section{The method}
%{\it The method.---}
Potential models have been quite successful in describing the properties of heavy quark bound states in vacuum (see {\it e.g.} \cite{Appelquist:1978aq}). Our key assumption is that at any temperature $T$, the interaction between heavy quarks can be  approximated by an instantaneous potential, $V(r, T)$, where $r$ is the radial coordinate.

We shall study bound states of heavy quarks by using  the non relativistic    Schr\"odinger equation 
\be\label{schrodinger}
\left( 2M_Q - \frac{\nabla^2}{M_Q} + V(r, T) \right) \psi_i = E_i  \psi_i  \,,
\ee
where $M_Q$ is constituent  mass of the heavy quark, $\psi_i$ and $ E_i$ represent the wave function and the mass of the corresponding $Q\bar Q$ state, respectively.
Comparing the results of this analysis with those obtained by the   Salpeter  equation for the $S-$wave states 
\be\label{salpeter}
\left(2\sqrt{M_Q^2 - \nabla^2}  + V(r,T) \right)\psi_i = M_i\psi_i \,,
\ee
we  find excellent agreement between the outcomes of the two methods, meaning that for these mesonic states relativistic corrections are small.

There are some controversies about how to extract the potential $V(r,T)$   from the color-singlet free-energy $F_1(r,T)$  (see  \cite{Wong:2004zr,Mocsy:2007jz,Satz:2008zc}). In principle, in order to obtain the internal energy $U_1$ one has to 
subtract the entropy contribution from the free-energy
\begin{equation}\label{potU}
U_1\,=\,F_1-T\frac{{\rm \partial} F_1}{{\rm \partial} T} \,.
\label{U1}
\end{equation}

However, this subtraction  procedure has been put to question, one of the reasons being that  around the critical temperature,  the potential $U_1$  is more attractive than the potential at zero temperature. 
In~\cite{Satz:2008zc} this effect has been interpreted as coming from an additional interaction between the constituents in the medium.  Below $T_c$ this effect should be due to hadrons and could be related to the  string ``flip-flop" interaction~\cite{Miyazawa:1979vx}; above $T_c$ it should be related to the antiscreening properties of QCD. In both cases one can consider that at finite temperature the static sources $Q$ and $\bar Q$ polarize the medium and a cloud of constituents forms around them. At short distances the main contribution to the potential energy comes from the $Q\bar Q$ direct interaction, which does not change with temperature. At intermediate distances the polarized clouds surrounding the two heavy quarks overlap and the interaction between the constituents of the clouds makes the interaction potential deeper than the potential at zero temperature. This indirect interaction leads to a change of the interaction strength at finite temperature and is extremely strong close to $T_c$. 
Some authors refer to this behavior of the internal energy as the ``over-shooting" problem and propose to address it by defining 
$V(r,T) = x U_1(r,T) + (1-x) F_1(r,T), 
$ for $0 \leq x\leq 1$, see Ref.~\cite{reviews}.  In agreement with Ref.~\cite{Satz:2008zc} we shall use the internal energy in  Eq.~(\ref{U1}) as the interaction potential. As we will see, with this potential we obtain dissociation temperatures for charmonia consistent with lattice results obtained with the Maximum Entropy Method~\cite{Asakawa:2003re}. Employing the free-energy as interaction potential gives much smaller dissociation temperatures. Therefore, in the following we shall assume that $x=1$; results  obtained with $x=0$ are shown for comparison. 

Notice that at large distances the two heavy quarks are screened and they do not interact: in this limit the internal energy can be interpreted as a contribution to the constituent mass of the quark. For this reason it can be useful to define $\bar U_1(r,T)=U_1(r,T)-U_1(\infty,T)$.

At short distances the static quark-antiquark  free-energy is normalized in such a way that it reproduces the free-energy at zero temperature
\be \label{cornell}
F(r, T = 0) =-\frac{\alpha}r +\sigma r \,,
\ee
where $\sigma $ is the string tension and $\alpha$ is the Coulomb coupling constant. The values of these two parameters can be phenomenologically fixed. Employing  the numerical values $\sigma = 0.16 $ GeV$^2$, $\alpha=0.2$ and with $m_c=1.28$ GeV one obtains $M_{J/\psi} \simeq 3.11$ GeV and  $M_{\chi} \simeq 3.43 $ GeV.

For non-vanishing temperatures we will consider the free-energy  proposed in~\cite{Dixit:1989vq} employing the formalism of the Debye-H\"uckel theory, with the parameterization of Ref.~\cite{Digal:2005ht}, 
\be\label{Fdixit}
F_1(r,T) = \frac{\sigma}{\mu} \left[ \frac{\Gamma(1/4)}{2^{3/2}\Gamma(3/4)} - \frac{\sqrt{x}}{2^{3/4}\Gamma(3/4)}K_{1/4}(x^2 + \kappa\, x^4)\right]-\frac{\alpha}{r}(\exp(-x)+x) \,, 
\ee
where $K_{1/4}$ is the modified Bessel function, $x=\mu \,r$, while the functions  $\mu \equiv \mu(T)$ and $\kappa \equiv \kappa(T)$ are determined by fitting the lattice data.  Once these two functions are fixed, this parameterization  of the free-energy is in excellent agreement with lattice data for $T \le 2\, T_c$. 

\begin{figure}[!th]
\includegraphics[width=3.in,angle=-0]{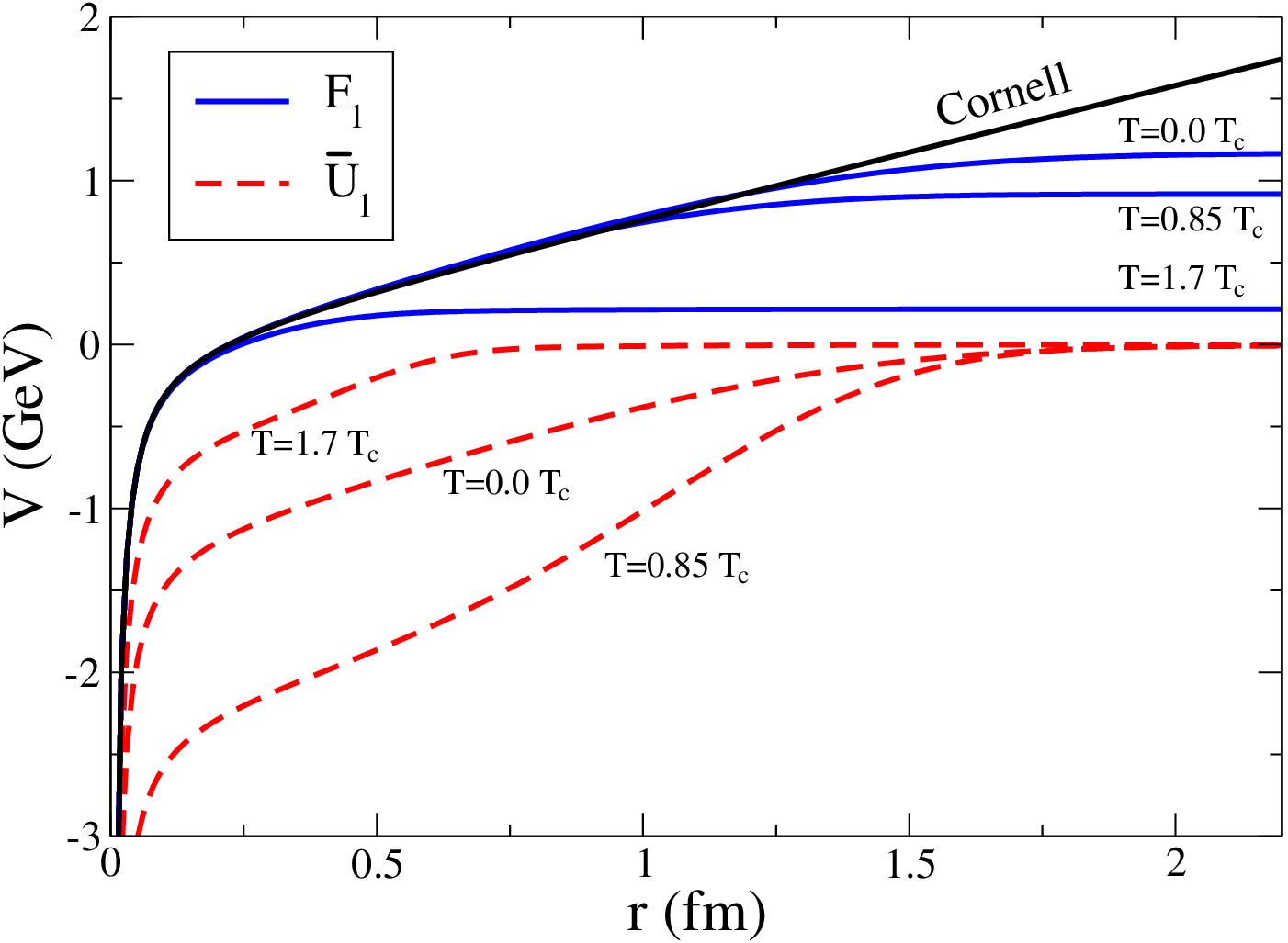}
\caption{(color online) Full black line corresponds to the Cornell potential, Eq.(\ref{cornell}). Full blue lines correspond to $F_{1}$ and dashed red lines  correspond to $\bar U_{1}$,  obtained with the  parametrization   in Eq.~(\ref{Fdixit}), for temperatures $T=0$, $T=0.85\, T_c$ and $T=1.7\, T_c$. Notice that $F_{1}$  and $\bar U_{1}$ have been normalized differently at large distances and that at vanishing temperature these two quantities only differ by a constant shift.}
\label{potentials}
\end{figure}

In Fig.~\ref{potentials} we compare the free-energy with the internal energy. 
When the free-energy is  used as a potential, it results in an interaction that always decreases its strength with increasing temperature. Moreover the only  effect that is present is the  screening of the interaction at large distances. 
On the other hand $\bar U_1$ is more attractive close to $T_c$ than at $T=0$, the representing curve in Fig.~\ref{potentials} is for $T=0.85 T_c$. Around $T_c$ the internal energy is extremely attractive; further increasing the temperature it eventually becomes weaker than in vacuum, as for the case of  $T=1.7\, T_c$ reported in Fig.~\ref{potentials}.

\section{ Masses and Decay widths}
%{\it Masses and Decay widths.---}
In order to determine the decay widths it is necessary to compute the wave function and the mass of the charmonia states. For $l=0$ states, they are determined solving both the Scrh\"odinger equation and the Salpeter equation through the Multhopp method~\cite{Colangelo:1990rv}  while for $P-$wave states we only solve the Schr\"odinger equation. 
For a potential with a singularity in $r=0$,  the wave function solution of the Salpeter equation is not well behaved at the origin and therefore we will  cut-off the short distance behavior of  the potential.  This procedure is analogous to the one employed in Ref.~\cite{Giannuzzi:2008pv} in vacuum. In particular, we introduce a cutoff radius, $r_{min}$, such that  for $r<r_{min}$ the potential is constant and equal to $V(r_{min})$. According to Ref.~\cite{Cea:1986bj}, we fix $r_{min} = \frac{4 \pi}{3 M}$. In all cases we find that the typical radius of quarkonia states is much larger than $r_{min}$, meaning that our results are not very sensitive to the value of the cutoff. 

Various decay processes will be considered. Annihilation  processes of charmonia can be viewed as two-stage 
factorized process.  The first process consists in  $c \bar c$  annihilation into gauge bosons. The annihilation takes place at the characteristic distance $r$ of order $1/m_c$, so for a non-relativistic pair $r \to 0$ and the annihilation amplitude is proportional to the wave function at the origin. Then, the produced  gauge boson  decays   or fragments into leptons and  hadrons.  In vacuum it is assumed that the  inclusive probability of the latter process is equal to one. At finite temperature we will assume that such a factorization persists. Therefore, the width of   hadronic  and  leptonic decays will depend on temperature  exclusively through the radial wave function at the origin and  the mass of the meson~\cite{Appelquist:1978aq}:
\be\label{width1}
\Gamma_A \propto \Gamma_{\ell^+\ell^-}  \propto \Gamma(^3\!S_1 \to 3 g) \propto    \frac{|R_{s}(T,0)|^2}{M(T)^2}\,.
\ee
We also consider the radiative transitions $\chi_{c J} \to J/\Psi + \gamma$ from the $^3 P_J$ levels to the $^3S_1$ state with rate 
\be\label{width2}
\Gamma_B \propto \Gamma_{\chi_{cJ} \to J/\Psi + \gamma}  \propto (2 J +1)  |I_{\rm PS}|^2\,,
\ee
where $ I_{\rm PS}$ is the  overlap integral between the radial wave functions of the corresponding states.  

\subsection{ A toy model for screening}
\begin{figure}[thdp]
\includegraphics[width=3.in,angle=-0]{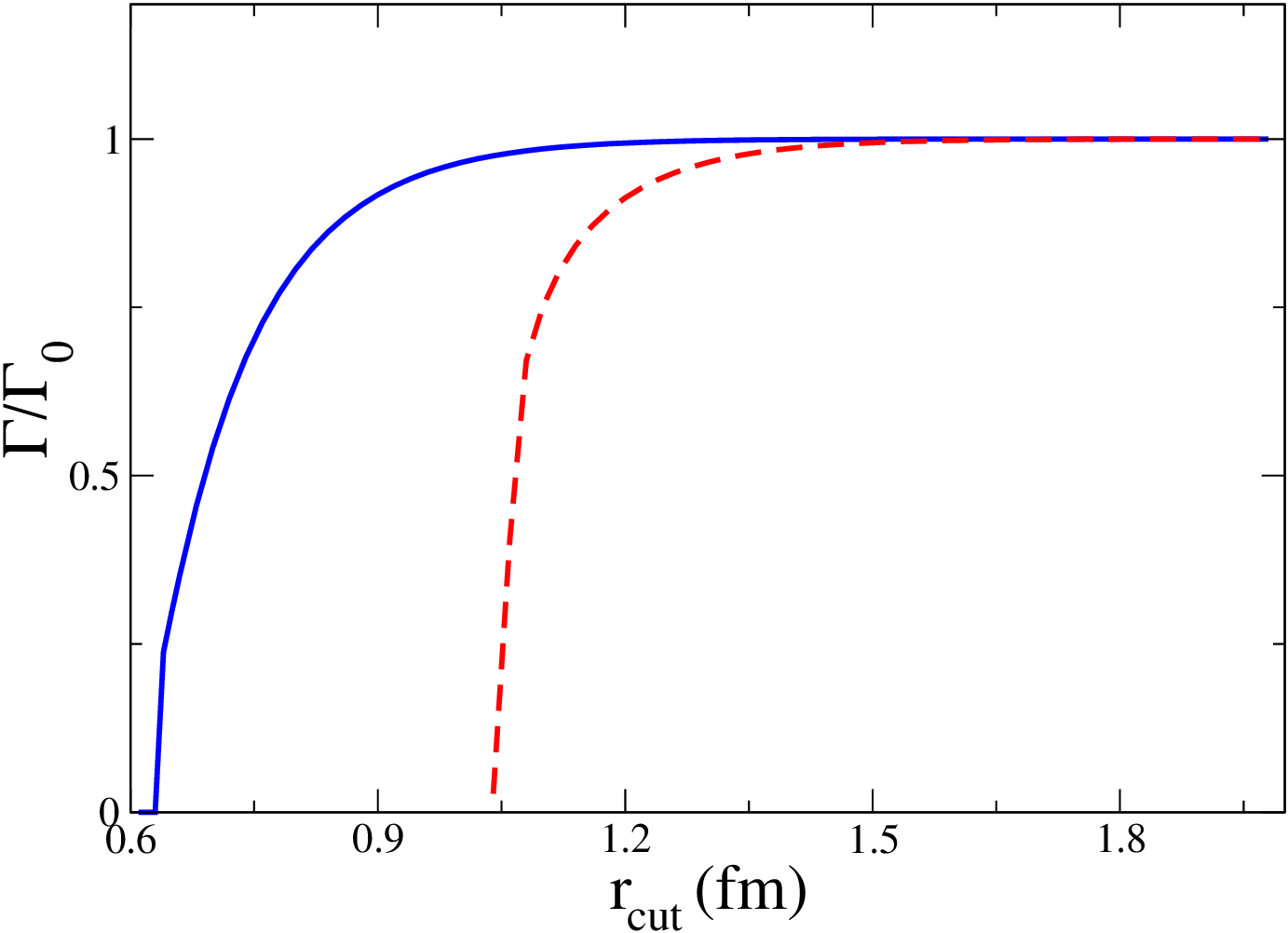}
\caption{(color online) Charmonia decay widths as a function  of $r_{\rm cut}$, obtained feeding the potential of Eq.~(\ref{eq-f-rcut}) in the Schr\"odinger equation (\ref{schrodinger}). Full blue line, decay width of Eq.~(\ref{width1}); dashed red line, radiative decay width of Eq.~(\ref{width2}).  The widths have been normalized to the  values obtained for $r_{\rm cut} \to \infty$, corresponding to the widths in vacuum.} \label{radiative-chic}
\end{figure}

We first introduce a toy model for studying the dependence of the decay widths on the screening length. It is based on the assumption that at short distance the interaction potential is given by the Cornell potential and is constant for $r > r_{\rm cut}$:
\be\label{eq-f-rcut}
V(r,r_{\rm cut}) = \left\{  \begin{array}{lc}
-\frac{\alpha}r +\sigma r & {\rm for }\,\, r< r_{\rm cut}\\ 
-\frac{\alpha}{r_{\rm cut}} +\sigma r_{\rm cut} & {\rm for }\,\, r> r_{\rm cut}\end{array} \right. \,.
\ee
This procedure roughly describes what happens for non-vanishing temperatures when one uses $F_1(r,T)$ as a potential, see Fig.\ref{potentials}. 
We find that the $\chi_c$ dissociates when $r_{\rm cut} \simeq 1.04$ fm, whereas the $J/\psi$ dissociates at  $r_{\rm cut} \simeq 0.64$ fm. The decay widths for the processes in Eqs.~(\ref{width1}) and (\ref{width2}) are plotted in Fig.~\ref{radiative-chic}   as a function of $r_{\rm cut}$.
The width of the process $\chi_{cJ} \to J/\psi + \gamma$  begins to decrease at $r_{\rm cut} \sim 1.2$  fm. At $r_{\rm cut} \simeq 1$ fm the width for this process is zero and the $\chi_c $ dissociates. The width for the processes in Eq.~(\ref{width1}) is  sensitive to smaller screening length, of less than $1$ fm. These results are in agreement  with the fact that these charmonia states have a radius of about $1$ fm and that the $J/\Psi$ is more compact than the $\chi_c$. However, the potential in Eq.~(\ref{eq-f-rcut}) does not take into account the contribution from the ``cloud-cloud" interaction~\cite{Satz:2008zc}. As we shall see,  it is not the screening of the interaction  that is important in characterizing these decay widths, but the antiscreening of the interaction. 

\subsection{Decay widths at different temperatures}
Now we show the results  for dissociation temperatures and  decay widths obtained  using as $Q\bar Q$ potential the  internal energy in Eq.~(\ref{U1}) and for comparison the  free-energy in Eq.~(\ref{Fdixit}).

We find that using the free-energy (\ref{Fdixit})  the $J/\psi$ dissociates at $1.2 \, T_c$, while   the $\chi_c$  dissociate at about  $ 0.95 \, T_c$. Using the internal energy as potential we find that the $J/\psi$ dissociates at approximately $1.8 \, T_c$, while the  $\chi_c$ dissociates at about $1.15 \, T_c$. We have estimated the dissociation temperatures from the binding energy of the pair.  Using the internal energy   in the wave equations, one obtains higher dissociation temperatures in agreement with the fact that $ U_1$ is more attractive than $F_1$. Employing the Salpeter equation we find a difference in the dissociation temperatures  of less than $ 5 \% $, in qualitative agreement  with results obtained in vacuum~\cite{Jacobs:1986gv}.

\begin{figure}[thdp]
\includegraphics[width=3.in,angle=-0]{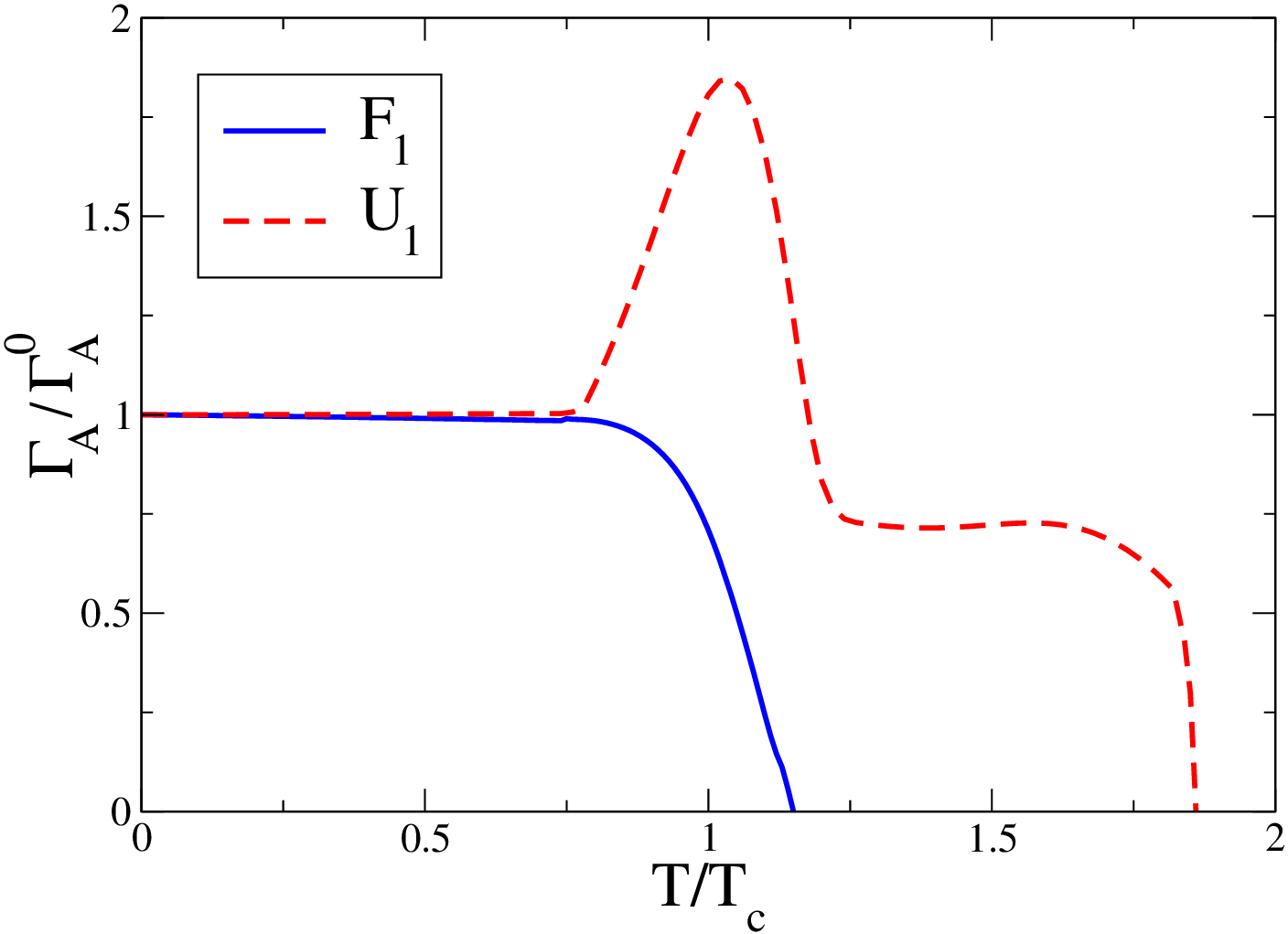} \hspace{1.5cm}
\includegraphics[width=3.in,angle=-0]{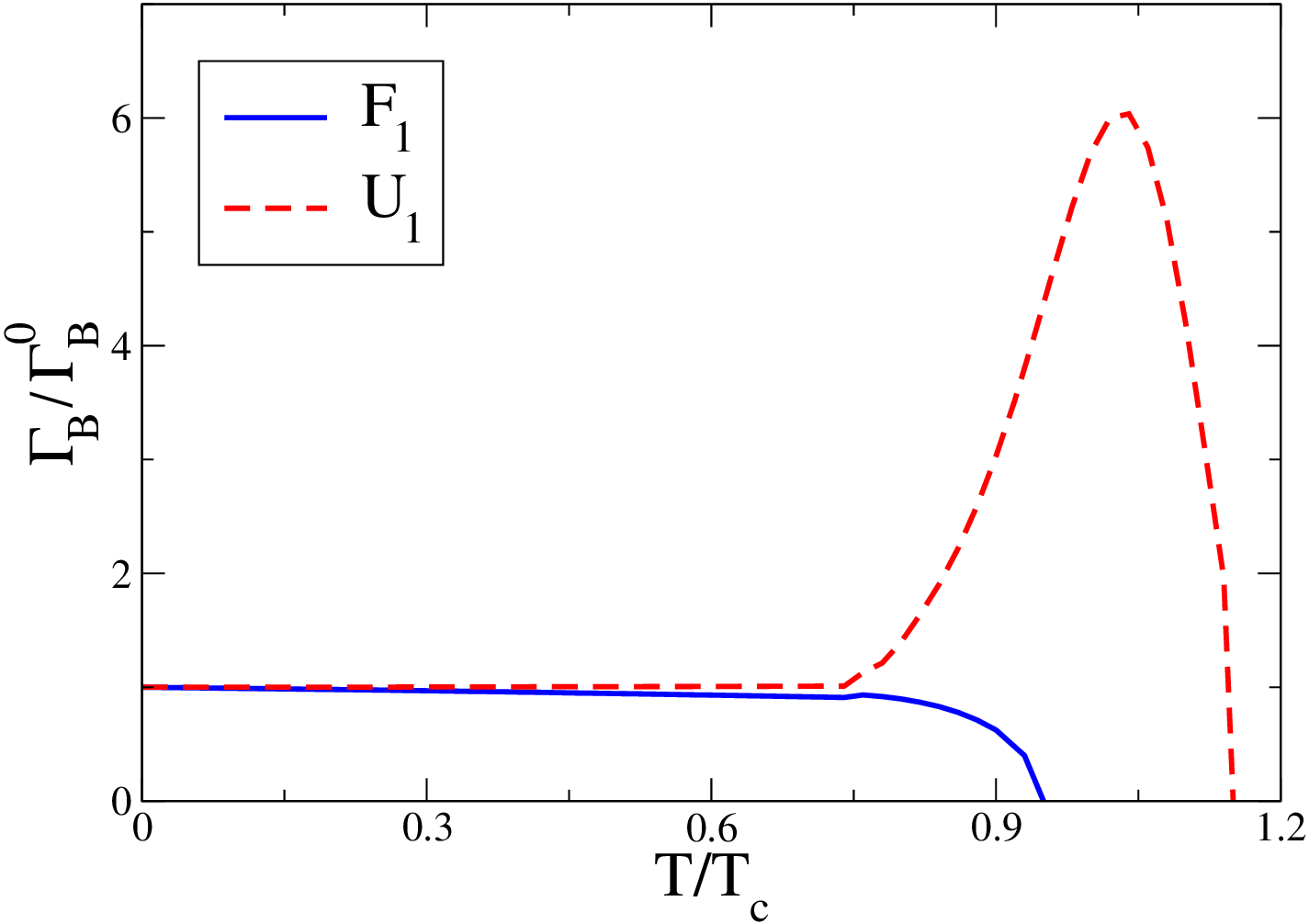}
\caption{(color online) Charmonia decay widths given in Eq.~(\ref{width1}) (left panel) and in Eq.~(\ref{width2}) (right panel). We have used  $F_{1}$, full blue lines, and $ U_{1}$, dashed red lines,  as potentials with the paramterization of  Eq.~(\ref{Fdixit}), and normalized the widths to the vacuum value.} \label{jpsi-origin}
\end{figure}

In Fig.~\ref{jpsi-origin} we report  the values of the decay width for the process in Eq.~(\ref{width1}), left panel,  and for the process in Eq.~(\ref{width2}), right panel,  as a function of the temperature. %The dashed red lines refer to the results obtained employing $ U_1$ as a potential, while the full blue lines are the results obtained using $F_1$ as potential.
Employing $F_1$ we find for both the decay processes considered here a behavior similar to the one reported in Fig.~(\ref{radiative-chic}). Therefore, one can interpret the variation of these decay widths with the temperature as due to the screening of the potential.
On the other hand,  when using  $ U_1$ we find that  across the transition region both decay widths become large. For the decays in Eq.~(\ref{width1}) the width is about a factor $2$ larger than in  vacuum. For the radiative decay in Eq.~(\ref{width2}), the ratio between the width in the thermal medium and in vacuum reaches a factor $6$ across the transition region. This effect is not due to the screening of the potential, but is rather connected with the ``cloud-cloud" interaction.  

In order to understand whether this effect is observable one should consider the evolution of the medium in collider experiments. For a rough estimate, let us suppose that the $\chi_c$ is produced in the early stage of the heavy-ion collision and then travels for about $4$ fm in the thermally equilibrated medium. If the temperature is sufficiently low, so that  the decay width is approximately the same as in vacuum,  one has that less than $1 \%$ of   $\chi_c$ decay. On the other hand, for temperatures close to  $ T_c$, one should observe that about $5\%$  of $\chi_c$ decay. Since for temperatures close to $T_c$ one can neglect interactions of charmonia  with in-medium hadrons~\cite{reviews, Matinyan:1998cb} and gluons~\cite{reviews} the radiative decay process  should give a sizable contribution to the total decay width of the $\chi_c$. As regard the inclusive width of the  $J/\Psi$, one should consider that  at temperatures below $T_c$,  the process in Eq.~(\ref{width1}) gives approximately the inclusive width. However, at larger temperatures the dominant contribution  should be  due to the interaction with in-medium partons~\cite{reviews}.

The largest uncertainty in our calculations resides
in the extraction of the potential from   
lattice QCD calculations and pertinent parametrizations to numerically evaluate
the  entropy.  In order to test the robustness of our results one can employ a different  parametrizations of the internal energy. Using the expression of  $U_1(r,T)$  reported in Ref.~\cite{Wong:2004zr}   we find approximately the same results reported here. In particular we find that in the crossover region the widths of the processes in Eqs.(\ref{width1}) and (\ref{width2})   are enhanced with respect to the vacuum values by approximately the same factors $2$ and $6$, respectively.    

The analysis reported in the present  paper can easily be extended to bottomia, moreover it  would be of some  interest to include   $D$ mesons.

\begin{acknowledgments}
We would like to thank N.~Armesto, P.~Colangelo, C.~Manuel, C.~A.~Salgado and J.~Soto for constructive comments and discussion.
The work of MM was supported by the Spanish grant
FPA2007-60275 and FPA2008-03918-E/INFN.

\end{acknowledgments}

%{\it Acknowledgment.---}

\end{document}